# Beyond “AI-Generated” Labels: Watermarking, Co-Creation, and Conflation of AI-Generation with Disinformation

Federico Germani[1,2*] 0000-0002-5604-0437

Giovanni Spitale[1] 0000-0002-6812-0979

[1]Information, Technology, and Experimental Ethics Lab, Institute of Biomedical Ethics and History of Medicine, University of Zurich, Switzerland

[2]Institute for Data Science and Artificial Intelligence, Boğaziçi University, Türkiye

*Corresponding author. Email: *federico.germani@bogazici.edu.tr*

## Abstract

Watermarking is often presented as a straightforward solution for distinguishing AI-generated from human-generated content, enabling platforms and regulators to trace synthetic content and detect AI-generated outputs at scale. This paper examines whether such mechanisms meaningfully address the epistemic and ethical challenges that arise in domains where the central concern is not the automation of content production, but the accuracy, intent, and deceptive potential of messages. We argue that extending watermark-based approaches to these settings is conceptually and practically misguided. Invisible watermarking encodes only model origin; when operationalized into visible “AI-generated” labels, it reduces complex creative processes to a misleading binary and provides no information about truthfulness. Such labels may stigmatize legitimate uses of generative tools while encouraging misplaced trust in unmarked content. Here we propose an alternative approach centered on process transparency and information literacy. We argue that these measures address the epistemic and ethical dimensions of AI-generated disinformation more effectively than visible watermark labels, reframing authorship as a transparent human practice rather than a binary indicator of machine involvement.

## Keywords

AI authorship; Watermarking; AI labels; Human–AI collaboration; Generative AI; Authorship transparency; AI policy; Process-based disclosure; Responsibility in AI use; Ethical AI tools; AI detection limitations; AI governance; Provenance

---

## The promises and technical limitations of watermarks

A watermark, in the context of generative AI, is a deliberately embedded signal — such as subtle statistical patterns in text, minute pixel or brightness adjustments in images, or inaudible modulations in sound — that acts as a digital signature linking content to its model of origin. The idea is to insert a persistent and ideally imperceptible trace into an output, hoping to make AI-generated material identifiable as such. Watermarks are technically interesting because they offer a mechanism to distinguish synthetic content from human-generated material at scale.

Distinguishing humans from machines is not a new challenge. A range of earlier systems have pursued similar goals in different domains, for example CAPTCHA-style tests designed to distinguish human users from automated bots (von Ahn et al. 2003; Foltýnek et al. 2019). While these systems serve different purposes, such as securing online infrastructures, they all share the common goal of identifying signals that can help institutions infer whether an action or interaction originates from human or automated processes. Now, watermarking for generative AI can be understood as a recent extension of this broader class of technologies; furthermore, it is important to mention that watermarking is situated within a broader ecosystem of provenance mechanisms designed to document the origin and transformation of digital content; these approaches include cryptographically signed metadata, content credentials frameworks, or perceptual hashing (Collomosse and Parsons 2024); in environments where automated systems can generate large volumes of text, images, or audio — such as spam pipelines, bots, or disinformation content farms — the prospect of reliably tagging AI outputs promises an efficient way to detect, filter, or slow down automated activity.

In what follows, we distinguish between invisible watermarking and visible AI labels. By invisible watermarking we mean machine-readable signals embedded into AI outputs at generation time, designed to encode model provenance for detection or forensic purposes. By visible AI labels we mean human-facing indicators such as “AI-generated” tags, which may or may not be derived from watermarking and are intended to inform audiences. Our critique focuses on two related issues: first, the way provenance signals from invisible watermarking can be operationalized into visible “AI-generated” labels for individual, co-created outputs, as if model origin mapped straightforwardly onto authorship; and second, the way such labels are then treated as proxies for the accuracy, reliability, or deceptive potential of content.

Now, in theory, a watermark should survive common edits and transformations while remaining detectable by specialized algorithms. One could argue that if AI outputs carried a durable, machine-readable signature, online platforms could automatically route, label, or limit their circulation; forensic analysts could attribute deepfakes to specific models; and academic journals could use such signals to flag undisclosed AI assistance in submissions. Watermarks would act as a means of restoring transparency and traceability in an information ecosystem increasingly shaped by synthetic content. And because of this, there is a vast amount of literature on identifying strategies to develop and deploy watermarks for synthetic text, audio, and video.(Zhao et al. 2023; Fu et al. 2024; Xu et al. 2025; Qu et al.)

That said, insofar the practical limitations are severe, across different output types. Empirical studies reveal that watermarks can be stripped or corrupted with minimal effort — through paraphrasing, resizing, noise injection, or regeneration attacks that erase up to 99% of the signal.(Zhao et al. 2025; Fu et al. 2025) Researchers have also shown that watermarks can be forged, falsely marking genuine human work as synthetic.(Saberi et al. 2023; Dong et al. 2025) The absence of shared standards compounds the issue: OpenAI, DeepMind, and Meta each employ distinct and incompatible watermarking schemes, (Gowal et al. 2025) while open-source or other models may include none at all. As a result, detection remains fragmented, inconsistent, and easily circumvented. Watermarks are thus arguably fragile, spoofable, and incomplete; and, most importantly for the purposes of transparency and traceability, blind to process, as they cannot capture degrees of human intervention, the role a tool played, or the intent behind its use.

# Watermarks and the problem of co-creation and authorship

Generative models are, at their core, instruments – extensions of human reasoning and imagination that operate within the boundaries of prompts, constraints, and iterative feedback. We could say, they are leverages of human intentions.(Germani et al. 2024a) They do not possess their own intentions or ethical discernment, and are not accountable per se. The human author remains the locus of authorship, the one who conceives, directs, evaluates, and assumes responsibility for meaning. This distinction between tool and author is not semantic but foundational.(Ginsburg and Budiardjo 2019) Authorship requires intention, an understanding of purpose, an ability to justify claims, and a willingness to be held accountable for outcomes. And it does not necessarily have to be the writer of a text; as Foucault puts it, the author is a construction, not a person, and that all authors are writers, but not all writers are authors.(Foucault 1979) AI tools, however sophisticated, operate without moral and epistemic agency.(Zafar 2025) Therefore, we argue that treating them as co-authors collapses a history of creative and scholarly practice in which tools have always mediated, but never replaced, human intent.

We say this because recognizing AI systems as tools within human-directed creative processes clarifies the limits of applying watermark-based distinctions at the level of individual outputs. Watermarking is well suited for contexts where the primary goal is to identify large-scale automated activity — such as bot networks, spam pipelines, or high-volume synthetic content streams — and can certainly help platforms reduce noise and curb coordinated manipulation. But these strengths do not carry over to situations where people use generative systems as part of an ordinary creative or scholarly workflow. In these settings, labeling a single document or image as either “AI-generated” or “human-generated” presupposes clean authorship boundaries that the creative process itself does not respect.

We do not, for instance, watermark a historian’s prose because a human copy-editor refined it; label a photograph as “camera-generated” because optics and sensors captured photons reflecting off landscapes or faces; consider a dataset’s figures as “software-generated” because a data visualization tool was used to generate representations of numbers as histograms; or mark a document as “Microsoft-Word-generated” simply because it was written within a word processor that suggested phrasing or formatting options. These examples represent a class of human–tool interaction, in which the tool contributes no creative or conceptual input and therefore cannot meaningfully be treated as a co-creator. The instruments, in these examples, facilitate expression, but do not originate meaning. The creative act resides in the human capacity to frame a question, define a goal, and judge the adequacy of the outcome. Generative AI tools therefore differ from earlier tools in their generative ability, which introduces a new scale of suggestion, variation, and synthesis; still, even in this case, the decisive act (the selection, contextualization, and ethical justification of what is produced) remains human.

In this context, creativity could arguably be understood not just as the autonomous production of novel ideas by a machine, but as a process of exploration, selection, and interpretation guided by human judgment. Creativity in human–AI co-creation resides in the interaction between computational suggestion and human selection and framing, rather than in the independent agency of the system itself (Boden 2004; Walia 2019; Celis Bueno et al. 2025; Holzner et al. 2025). Thus here, co-creation (which in the context of generative AI, refers to the integrated act of meaning-making in which human cognitive, ethical, and aesthetic choices guide and reshape the outputs of computational systems) becomes possible: not because the system has agency, but because it can propose content that the human author evaluates, integrates, reshapes, or discards. It is not a symmetrical partnership but a dynamic interplay in which machines contribute exploratory possibilities (prompted by humans), while humans supply framing, selection, accountability, and intent. The power of co-creation lies precisely in the dialectic of an iterative, reflective process through which humans curate computational and statistical possibilities into purposeful and meaningful communication — although in alignment with the Lovelace effect, which posits that humans tend to perceive and project creativity onto AI systems and their outputs; in this sense human observers may nonetheless experience these outputs as surprising or creative, often supporting human creative processes, even though the models merely follow instructions and lack the capacity to originate or generate meaning and understanding (Natale and Henrickson 2024). In this sense, human–AI co-creation can also be seen as extending longer traditions of collaborative creative practice in which authorship emerges through distributed and iterative contributions among multiple actors. (Cizek et al. 2022)

It is worth emphasizing why these arguments matter philosophically. Watermark-based detection assumes that meaningful content emerges from discrete, separable acts of authorship, when in reality creative production is processual and relational. Of note, this is not the case fully automated outputs such as those generated by bots, which involve no creative process and therefore raise an entirely different set of questions, which can be addressed through watermarking. In general, current AI tools enter this process as collaborators in form, not in intention — they extend the human act of creation without replacing its origin in consciousness, and they carry the same accountability we ascribe to a word processor or a camera, namely, none. We contend that any attempt to mark AI participation as an independent authorship event misunderstands what co-creation is: a continuum of interaction, revision, and interpretation in which human agency remains the unifying thread.

Watermarking's limitations become clearer when we consider how they intersect with the lived reality of creative and professional practice. AI is often used as a tool or partner rather than an independent author, which blurs the lines of authorship in ways a watermark cannot readily capture. For instance, an artist might use an AI image generator for an initial concept, then heavily edit or paint over it by hand; the final artwork is neither purely AI nor purely human-made. Yet an intact invisible watermark might still cause the work to be classified or labeled as "AI-generated," effectively ignoring the human contribution. Conversely, if editing is substantial enough to erase the watermark, a detector might treat the piece as 100% human-made, erasing the AI's role. In both cases, the degree of each party's involvement is lost. Of course, watermarking is not designed to measure degrees of human involvement; its primary purpose is to signal model provenance at the moment of generation, often for domain-specific governance such as deepfake detection or forensic verification. The challenge we highlight is that when watermarking is extended beyond these infrastructural use cases and applied to individual outputs, it becomes misaligned with the relational and iterative nature of human–AI collaboration. In line with this, researchers from a recent human-computer interaction study note that current watermarking methods “may not capture the full complexity of co-creative workflows.”(He and Do) The authors argument with two examples: 1) Meta’s Stable Signature watermark, which remains regardless of subsequent modifications to an AI-generated image – no matter how much a human artist fine-tunes the image afterward, the embedded mark continues to attribute the image to the AI model just as if no human had intervened; and 2) Google’s SynthID watermark for text, which shows moderate resilience to minor edits, yet its detectability drops considerably once the text is extensively rewritten or translated.(He and Do) In practice this means a lightly edited AI draft might still trigger detection (labeling the whole text AI-generated), whereas a heavily rewritten passage might evade detection entirely. Watermark signals do not degrade in proportion to human edits – they tend to function in a binary all-or-nothing way, which is poorly suited to reflecting “degrees” of AI assistance.(Rastogi and Pruthi 2024; Harel-Canada et al. 2025; Cheng et al. 2025)

Now, let’s see how watermarks would apply in different examples, each illustrating how co-creation complicates attempts to label or separate human and AI contributions.
First example: A research group drafts a paper. A model helps restructure the introduction and tightens the prose in the discussion. The argument, the evidence, the claims, and the accountability remain the authors’. A watermark on the final version of the manuscript would not tell readers what matters: that AI assisted in copy-editing and paragraph flow, that factual assertions were independently checked, and that no text was fabricated or hallucinated.
Second example: A newsroom uses speech-to-text to transcribe interviews, a summarizer for background documents, and a generator to suggest alternative headlines. Editors choose, rewrite, fact-check, and publish. What would a watermark mean here? Which layer would be labeled — the transcript, the notes, the headlines, the article?
Third example: An artist prototypes concepts with a diffusion model, then paints over prints and composites elements by hand. A durable image watermark would “freeze” a first pass of the workflow and misattribute authorship of the final piece; an easily erased mark would do nothing.

At this point we could argue that the problem of watermarks in the context of co-creation and authorship is not merely theoretical. *Hypnocracy*,(Xun 2024) a book revealed to have been co-written by AI, illustrates this case. Italian author Andrea Colamedici initially published the book under the fake name Jianwei Xun, to conduct a philosophical experiment in AI-human co-creation and authorship. After the process was publicly revealed, Colamedici explained that some of the book’s content was AI-generated (produced via ChatGPT and Claude), but that he curated, edited, and interwove the AI passages with his own writing. He argued that it's impossible to clearly say who wrote what, as the writing itself would not have been possible without the machines, as well as without the author himself. That is why the co-creation of AIs and humans represents, according to Colamedici, a new creative process that generates new knowledge.(Colamedici 2025) In *Hypnocracy*, entire paragraphs were a fusion of human and artificial ideas blended together, machine outputs and the author's refinements.(Colamedici 2025) Before examining this case, it is worth noting that Colamedici himself brought the issue of human–AI co-creation into public debate with *Hypnocracy*. Yet the phenomenon he described is far from unique. Many writers, scholars, illustrators, musicians, and video editors have already been engaging in similar practices, using AI tools as extensions of their creative process. In fact, we, the authors of this paper, are doing precisely the same as we write and refine these arguments through AI-assisted drafting and revision. Thus, Colamedici's book serves as a philosophical lens through which a much broader and ongoing shift in creative production can be discussed.

If one had a (functional) detector for AI text, what would they conclude? It certainly would not meaningfully quantify or represent the balance of human and AI input – it would either flag certain sections or (depending on the watermark robustness) possibly label the whole work as human if enough editing broke the patterns. We therefore argue that watermarking offers no mechanism to grade or quantify partial AI involvement. As Colamedici's experiment demonstrated, even seasoned experts and critics failed to recognize the AI's contribution to the final text (though, to be fair, Gino Tocchetti did)(Tocchetti 2025) in line with research results showing the inability of people to distinguish between AI-generated text and organic text (Spitale et al. 2023) — and a simplistic AI label would fail to illuminate the subtle "co-authorship" at play. Philosophically, this raises the question: if a human mind guides the structure and integrates the ideas, but an AI generates chunks of text, is the result "AI-generated" or human-generated? The answer defies a binary category, and watermarking by itself cannot resolve that ambiguity.

Some researchers have pointed out that users' perceptions of ownership and credit in co-created works vary widely.(He and Do) As we do in this paper, many authors see generative AI as just a tool, and thus still consider themselves the primary author, even if the AI contributed chunks of text or helped improving language flow. In some circumstances, they may be reluctant to credit the AI at all – indeed, disclosing AI involvement can carry stigma.(Dergaa et al. 2023) Studies have found that audiences often perceive AI-assisted art or writing as less authentic or lower-value; Messer argued that AI usage can decrease the perceived authenticity and value of artwork, which leads many authors and artists to avoid revealing that they used co-creation processes with AI.(Messer 2024)

This discussion aligns with recent debates in academic publishing about the appropriate role of AI in scholarly writing. Porsdam Mann et al. (2024) have proposed an ethical framework for the responsible use of large language models in research, outlining three criteria: human vetting, substantial human contribution, and acknowledgement and transparency.(Porsdam Mann et al. 2024) Their approach situates AI use within long-standing norms of research integrity, distinguishing between *outcome goods* (the advancement of knowledge) and *process goods* (the fairness and accountability of how that knowledge is produced).(Porsdam Mann et al. 2024) These principles provide a practical counterpart to our philosophical claim: that watermarking misconstrues authorship by focusing on output detection rather than process integrity. In line with this reasoning, we argue that what defines responsible creation of text, images, or videos, is not the absence or presence of AI, but humans-in-the-loop, including human oversight, verification, and disclosure that ensure epistemic and ethical accountability.

Finally, we argue that watermarking and the visible labels that can derive from it invite strategic manipulation. The presence of a watermark could be forged to discredit authentic text or media; the absence of a watermark could be weaponized as presumed human authenticity even when content is synthetic.(Li et al. 2025) On the one hand, if watermarks become commonly used to label AI-generated content, people may start assuming unmarked content is real, which is not guaranteed (the "false negative" problem). On the other hand, people can deny reality by claiming something true is actually AI-generated. (Chesney and Citron 2019) We argue that public trust for information does not hinge on watermark presence; it depends on verifiable context, provenance, and process integrity. Trust emerges from transparency, consistent standards, and the credible accountability on the part of those who create and disseminate information.(Germani et al. 2024b; World Health Organization 2025) Without these human-centered mechanisms, we contend that even the most sophisticated watermarking systems only risk amplifying suspicion.

## Conflating synthetic with inaccurate

The concerns raised so far pertain to how watermarking interacts with human–AI co-creation and the difficulty of mapping binary labels onto distributed creative processes. A further, and distinct, issue emerges when the presence of a watermark is taken to signal something about the content itself; as if identifying a work as "AI-generated" provided information about its accuracy, reliability, or potential to mislead. This is not a limitation of watermarking as a technical mechanism, but of how it can be misused or misinterpreted in public and policy discourse. When watermarking is coupled with visible "AI-generated" labels and treated as an epistemic signal, it risks encouraging a conflation between the synthetic nature of a piece of content and the assessments of its truthfulness or intent. Conflating synthetic content with inaccurate or manipulative content, such as disinformation, poses serious conceptual and practical problems. Not all AI-generated material is false, harmful, or deceptive. The synthetic nature of a work says nothing about its epistemic accuracy or ethical intent. A 'fake' text can contain accurate, verifiable information, and in fact research shows it can even be more compelling than factual information written by humans.(Spitale et al. 2023) Rather than undermining trust, synthetic content, if transparently produced and responsibly verified, can advance public understanding.(Germani et al. 2024a) This insight is particularly relevant for public health communication, where accessibility and clarity are critical.(Purnat et al. 2023) The persuasive and stylistic strengths of generative AI models can help adapt complex information to diverse audiences, improve health literacy, and reduce conspiratorial thinking.(Dunn and Hazzard 2019; Nutbeam 2023; Fitzpatrick 2023; Costello et al. 2024; Sezgin and Kocaballi 2025) If AI-generated contents were automatically stigmatized by virtue of being 'synthetic' through visible watermark labels, communicators would lose valuable tools for enhancing public dialogue. As for text, synthetic visual or audiovisual media are not inherently manipulative. An AI-generated video might animate medical data for teaching purposes or visualize ethical dilemmas in medical training. Likewise synthetic images can simplify complex data and can be used for infographics. In general, synthetic content can be used for epistemically and pedagogically valuable uses that should be encouraged, not discouraged.

The problem with equating synthetic content to disinformation lies in collapsing form with intent. Disinformation is defined not by how content is created but by the intention to deceive. (Council of Europe 2017) A human can fabricate harmful lies, while an AI can produce transparent, accurate educational materials. Therefore, when public discourse treats all AI outputs as suspicious, the attention shifts away from the human actors' intentions, the verification processes in place, and the broader communicative context, which are the aspects that matter.(Germani et al. 2024a)

Moreover, arguably such conflation risks creating a chilling effect on innovation. Artists, educators, and communicators may hesitate to use generative tools ethically because audiences or regulators might misinterpret their use as deceptive. This could slow the integration of AI into socially beneficial applications, including accessibility tools, multilingual communication, and data visualization, beyond the benefits and results of co-creation itself. The epistemic and ethical evaluation of content should focus on truthfulness and intent, not on the technical origin of the media. Synthetic media are not inherently less trustworthy; their legitimacy depends on transparent processes of creation, human oversight, and verifiable accuracy. Recognizing this distinction is essential for designing mechanisms within fragmented governance structures that encourage beneficial uses of AI while effectively targeting genuinely harmful manipulative practices.

Furthermore, we argue, creating resilience against disinformation lies not in determining whether content is synthetic but in equipping people with the skills to evaluate the communicative strategies and epistemic claims embedded within it. The race to reliably distinguish AI-generated from human-generated content is, in practice, a lost cause since the quality of AI-generated text, images, and videos has been becoming increasingly high, making it impossible in the near future for people to distinguish AI-generated content from human-generated content. Even if perfect detection were possible through watermarking, it would still not answer the normative questions that matter: whether the message conveyed by such contents is accurate, whether it is manipulative, and whether the intentions are good. This is where information literacy and critical thinking skills becomes essential.(Redaelli et al. 2025) Rather than focusing on the medium of creation, information literacy approaches would teach individuals to identify misleading rhetorical tactics, assess evidence, scrutinize sources, and recognize emotional manipulation or epistemic red flags. Information-literacy interventions therefore shift the focus from what produced the message to what the message says, thus empowering individuals to evaluate claims regardless of their synthetic or human origin.

## Foreseeing objections and debating

A first objection one may have is that visible watermark labels deter deception. Deterrence for deception is desirable, but the cost is mislabeling legitimate co-creation and creating false confidence in unmarked content. The assumption that deterrence alone justifies labeling ignores the complex epistemic environment in which information circulates. Watermark labels may indeed dissuade careless misuse or low-effort deception, but they can also stigmatize ethical creative work and inadvertently convey false reassurance. When audiences think that ‘unmarked’ means ‘safe,’ they may relax critical scrutiny precisely where it is needed most. Conversely, authors and artists who use AI transparently may face undue suspicion or moral judgment.

A second objection might be that the public wants to know when AI is used. This is a legitimate concern, but what the public truly seeks is not a binary answer about AI involvement but clarity about what kind of involvement took place. Members of the public want to know whether a quote is real, whether a claim is verified, and whether human experts reviewed the final work. In reality, public understanding and trust depend on contextual information (what was generated, how it was verified, and by whom). A short, human-readable disclosure conveys this nuance far better than a visible watermark label. For example, in academic papers stating that AI assisted in editing and formatting, and that all content was reviewed by the authors, communicates accountability and transparency. Instead, visible watermark labels fail to inform or empower audiences, reducing a complex process to a meaningless binary label (for discussions on responsible practices for synthetic media, see Partnership on AI's work (Partnership on AI 2023); or how Meta handles labeling of AI-generated content and manipulated media (Bickert 2024))

## Conclusions

In this paper we argued that while invisible watermarking can play an important role in identifying fully automated outputs, such as those produced by bots, its usefulness does not extend to settings where humans and AI systems work together in the production of content; from a human-AI co-creation perspective, watermarking fails to accommodate the "blended" nature of creative workflows. Invisible watermarking can be useful for tracing the origin of fully automated outputs, but this strength does not translate to contexts in which human and AI contributions are interwoven. In this context, visible labeling of such content using "AI-generated" tags reduces a complex process to a misleading binary. There is a continuum of involvement, and often AI is just one part of the creative process. Watermarking and labels, as currently conceived, cannot indicate how much or in what way AI contributed. It also cannot capture intent — whether AI was used to brainstorm, draft, refine, or just automate low-level tasks. We also questioned the usefulness of watermarking from a philosophical standpoint; if we live in a time of pervasive human-AI co-creation, trying to neatly separate "AI content" from "human content" may be a fundamentally flawed attempt. Our critique echoes the point raised by Colamedici's *Hypnocracy* stunt,(Xun 2024) that the boundary between human and machine authorship can be so porous that any labeling scheme which treats them as separable will fail to do justice to reality. Our analysis highlights not only the conceptual and practical shortcomings of visible watermarking labels in the context of human–AI co-creation, but also the epistemic risks of conflating synthetic content with disinformation. Treating all AI-generated outputs as inherently deceptive collapses the distinction between form and intent; synthetic does not mean false. A generative model can produce accurate, transparent, and educational materials just as easily as a human can produce manipulative ones without using AI. We therefore argued that the problem lies not in the medium of creation, but in the ethics and accountability of its use. A useful and effective discussion of governance should therefore move beyond treating watermarking as a universal solution and instead differentiate between domains. At the infrastructure level, watermarking can support platforms in managing high volumes of automated content, where human-mediated scrutiny is impossible. Complementary technical approaches, such as perceptual hashing, may help document aspects of the production process and provide contextual information about how media were generated and modified over time. However, even these infrastructures cannot fully capture the interpretive and evaluative dimensions of human–AI co-creation. In contexts involving individual messages or creative outputs, meaningful transparency ultimately depends on human-centered disclosure that clarifies how and why AI was used and what forms of oversight and verification were applied. Such process transparency cannot scale to billions of objects and is not meant to; what it can do is prevent the collapse of "synthetic" into "deceptive" and support epistemic accountability in the assessment of claims and intentions. Finally, complementing this, we argued that information-literacy and critical thinking interventions are essential, since they empower individuals with the skills to assess accuracy, intent, and manipulative strategies of messages, regardless of whether a piece of content is generated with AI or not.